\documentclass[aps,twocolumn,nofootinbib, preprintnumbers, superscriptaddress]{revtex4}

\usepackage{amsmath, amssymb, slashed, braket, bm}
\usepackage{graphicx}
\usepackage{epstopdf}
\usepackage{float,appendix}
\usepackage[colorlinks=true,
            linkcolor=blue,
            urlcolor=blue,
            citecolor=green,
            bookmarks=true,
            bookmarksnumbered=true,
            breaklinks=true,
            pdfpagemode=FullScreen,
            pdfstartview=FitBH]{hyperref}
\usepackage{esint}
\usepackage[normalem]{ulem}
\usepackage{siunitx}
\usepackage{multirow}
\usepackage{soul}
\usepackage{xcolor}

\graphicspath{{figs/}} 

\usepackage{xcolor}
\definecolor{gesfpurple}{rgb}{0.47,0.19,0.42}

\definecolor{gesflanse}{rgb}{0.00,0.50,0.50}

\definecolor{gesfblue}{rgb}{0.08,0.42,0.76}

\definecolor{gesfred}{rgb}{1,0,0}

\definecolor{gesfwhite}{rgb}{1,1,1}

\definecolor{gesfblack}{rgb}{0,0,0}

\newcommand{\geqn}[1]{Eq.\,\hypersetup{linkcolor=blue}(\ref{#1})\hypersetup{linkcolor=blue}}
\newcommand{\gfig}[1]{{\hypersetup{linkcolor=violet}Fig.\,\ref{#1}\hypersetup{linkcolor=blue}}}

\newcommand{\prlsection}[2]{{\it\textbf{#1}{#2}}---}

\begin{document}

\title{Macroscopic Quantum Interference in Dark Matter Wave Scattering \\ with MICROSCOPE}

\author{Cheng-Tao Fu}
\affiliation{National Gravitation Laboratory, MOE Key Laboratory of Fundamental Physical Quantities Measurement,
and School of Physics, Huazhong University of Science and Technology,
Wuhan 430074, People’s Republic of China}

\author{Peng-Shun Luo}
\email{pluo2009@hust.edu.cn}
\affiliation{National Gravitation Laboratory, MOE Key Laboratory of Fundamental Physical Quantities Measurement,
and School of Physics, Huazhong University of Science and Technology,
Wuhan 430074, People’s Republic of China}

\author{Rui Luo}
\affiliation{National Gravitation Laboratory, MOE Key Laboratory of Fundamental Physical Quantities Measurement,
and School of Physics, Huazhong University of Science and Technology,
Wuhan 430074, People’s Republic of China}

\author{Jie Sheng}
\email{jie.sheng@ipmu.jp}
\affiliation{Kavli IPMU (WPI), UTIAS, University of Tokyo, Kashiwa, 277-8583, Japan}

\author{Chuan-Yang Xing}
\email{cyxing@upc.edu.cn}
\affiliation{College of Science, China University of Petroleum (East China), Qingdao 266580, China}

\begin{abstract}

Ultralight dark matter behaves as a coherent wave, yet its quantum interference effects of elastic scattering with multiple targets have remained unexplored. We show that the nested test masses of MICROSCOPE realize such an ``interferometer'' for dark-matter wave scattering. Amplitudes from the two concentric cylinders interfere and redistribute the induced force between them. This effect produces unique and  rotation-modulated signals set by the target geometry. Developing the theoretical framework and applying it to MICROSCOPE data, we obtain leading constraints on quadratic dark-matter--nucleon coupling for masses $10^{-3}$--$10^{-2}\,$eV, reaching cross sections of order $10^{-52}$\,cm$^2$.

\end{abstract}

\maketitle

\prlsection{Introduction}{.}
The nature of dark matter (DM) remains one of the central open questions in fundamental physics~\cite{Bertone:2004pz,Bauer:2017qwy,Cirelli:2024ssz}. A wide range of gravitational and cosmological observations has established its existence~\cite{Zwicky:1933gu,Rubin:1980zd,Clowe:2006eq,Planck:2018vyg}, yet its non-gravitational properties remain unknown. This has motivated a broad experimental program to search for DM over a wide range of masses and coupling strengths~\cite{Battaglieri:2017aum,Billard:2021uyg}. In particular, ultralight DM~\cite{Marsh:2015xka,Hui:2016ltb,Ferreira:2020fam,Antypas:2022asj,Kimball:2023vxk,Eberhardt:2025caq} has attracted considerable attention because it gives rise to qualitatively new detection strategies~\cite{Eberhardt:2025caq} beyond conventional direct-detection experiments based on scattering-induced excitations~\cite{Goodman:1984dc,Lewin:1995rx}.

A distinctive feature of ultralight DM is its wave nature. Owing to its tiny mass and large occupation number, the DM background is well described as a coherent wave~\cite{Preskill:1982cy,Abbott:1982af,Dine:1982ah,Hui:2016ltb,Kimball:2023vxk}. Long-wavelength DM quadratically coupled to nucleons can scatter coherently from all particles in a macroscopic target, greatly enhancing the total scattering cross section~\cite{Fukuda:2018omk,Akhmedov:2018wlf,Afek:2021vjy,Fukuda:2021drn,Day:2023mkb,Luo:2024ocg,Matsumoto:2025rcz,Gan:2025nlu,Liu:2025jlx}. The resulting continuous momentum transfer can then induce observable forces or accelerations on the target. Existing proposals have shown that this coherence can strongly enhance the response of precision acceleration sensors, making torsion-balance tests of the equivalence principle (EP)~\cite{Schlamminger:2007ht,Wagner:2012ui} the most powerful probes of ultralight DM--nucleon interactions~\cite{Luo:2024ocg,Matsumoto:2025rcz,Gan:2025nlu,Liu:2025jlx}.

The MICROSCOPE space experiment~\cite{Touboul:2017grn,MICROSCOPE:2019jix,MICROSCOPE:2022doy}, another precision test of the weak equivalence principle (WEP), features superior sensitivity and a larger target size, and should therefore achieve even greater sensitivity to ultralight DM. However, its nested multi-component structure differs qualitatively from the single macroscopic target usually assumed in theoretical treatments~\cite{Luo:2024ocg,Matsumoto:2025rcz,Gan:2025nlu,Liu:2025jlx}. When the sizes and separations of the target components become comparable to the DM wavelength, the standard description of DM scattering from a single object is no longer sufficient. In this regime, the DM wave exhibits not only coherent enhancement but also interference effects analogous to double-slit optics. This phenomenon has not yet been explicitly explored either theoretically or experimentally.

\begin{figure}[t]
    \centering
    \includegraphics[width=1\linewidth]{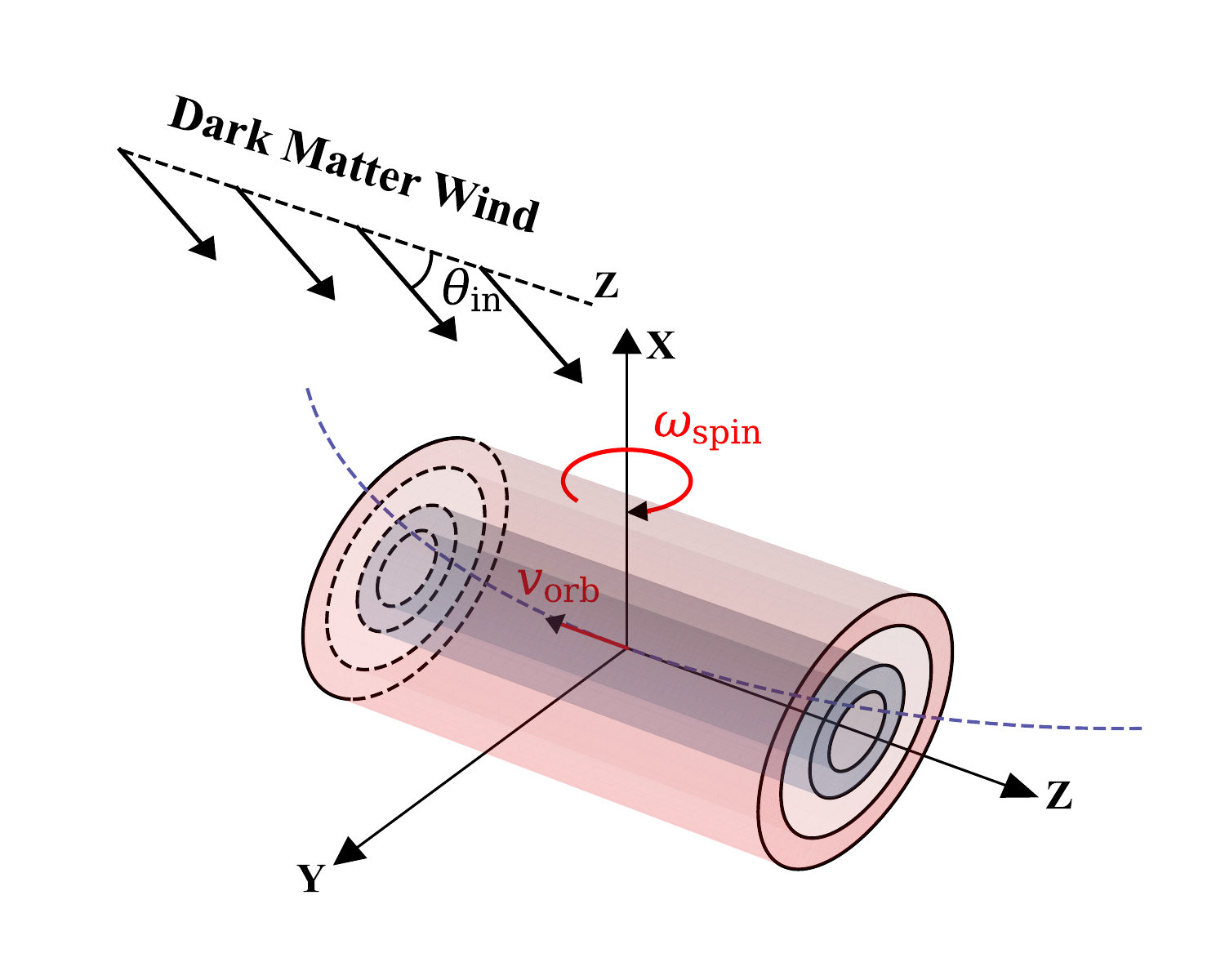}
    \caption{Schematic illustration of the MICROSCOPE test masses, showing their orbital motion and spin about the axis normal to the orbital plane. The DM wave wind is incident at an angle $\theta_\mathrm{in}$ relative to the sensitive $z$-axis.}
    \label{fig:cartoon}
\end{figure}

In this Letter, we show that interference in DM wave scattering can produce distinctive and observable signatures in precision acceleration measurements. We develop a formalism that captures the interference between scattering amplitudes from different macroscopic targets and identify the regime in which it becomes important. Applying this framework to MICROSCOPE as shown in \gfig{fig:cartoon}, we find that this experiment is sensitive to the predicted signal and can set stronger limits on DM--nucleon interactions in the mass range of $10^{-3}\!-\!10^{-2}\,\mathrm{eV}$.
Our results demonstrate that precision tests of fundamental physics can also serve as laboratories for wave interference effects in DM scattering.

\prlsection{Dark Force Redistribution from Quantum Interference}{.}
Elastic scattering is one of the most ubiquitous and minimal forms of interaction between fundamental particles. Assuming that there exists a quadratic interaction~\cite{Hees:2018fpg,Bouley:2022eer,Banerjee:2022sqg,VanTilburg:2024xib,Burrage:2026loe} between DM $\chi$ and nucleons ($N = n, p$), $\mathcal{L} \sim \bar m_N \chi^2 \bar N N /\Lambda^2$, elastic scattering can occur, with scattering cross section $\sigma_{\chi N}$. For ultralight DM of velocity $v_\chi$ and mass $m_\chi$ scattering off a nucleon, the momentum transfer amplitude is of order
$q \simeq m_\chi v_\chi = k$ where $k$ is the magnitude of DM incident momentum.
These interactions cause DM to scatter continuously off a macroscopic target, thereby inducing a detectable force on it.

We begin with a single macroscopic target $1$ with potential $V_1 \propto \sum_i \delta({\bf x} - {\bf x}_i)$ where each $\delta$-function represents the contribution from an individual nucleon~\cite{Fukuda:2018omk}. On macroscopic scales, it can be coarse-grained into a contact potential with shape of the target. In the first Born approximation, DM with local density $\rho_\chi = 0.4\,$GeV$/$cm$^3$ and flux $j_{\rm in} = \rho_\chi v_\chi/m_\chi$  would generate a force on target along its incident $z$-axis as
\begin{equation}
F_{1,z} (k)
=
j_{\rm in} k
\int d\Omega\,(1-\cos\theta)\,|f_1({\bf q})|^2.
\label{force_single}
\end{equation}
Suppose that, after scattering, the DM momentum changes from ${\bf k}$ to a final momentum ${\bf k}'$, with the angle between them given by $\theta$ and the momentum transfer $\mathbf q \equiv \mathbf k-\mathbf k'$. The scattering amplitude for this process is then
\begin{equation}
f_1(\mathbf{q})
=
-\frac{m_\chi}{2\pi}
\langle \mathbf{k}'|V_1|\mathbf{k}\rangle .
\end{equation}
For a target composed of $N$ nuclei located at positions $\mathbf{r}_i$, the amplitude can be calculated as
\begin{equation}
|f_1(\mathbf q)|^2
=
|f_{\chi N}(\mathbf q)|^2 \sum_{i,j} e^{i\mathbf q\cdot \Delta \mathbf r_{ij}}\,,
\label{phase_sum}
\end{equation}
with $|f_{\chi N}(\mathbf q)|^2 = d \sigma_{\chi N}/d \Omega$ and $\Delta \mathbf r_{ij}$ the displacement between different pairs of nuclei.
This expression includes the coherent enhancement of the scattering between ultralight DM and macroscopic object. If $\mathbf q\cdot \Delta \mathbf r_{ij} \ll 1$, the phase summation in \geqn{phase_sum}, which is defined as the scattering form factor, would contribute an enhancement factor of $N^2$.

When multiple targets are present, and their sizes and separations are comparable to the DM wavelength, DM can scatter coherently from all particles within each target. In addition, interference effects between different targets can be non-negligible and give rise to distinctive force signatures in DM detection.
We now generalize to two scatterers with centers at $\mathbf R_1$ and $\mathbf R_2$.
The total potential is
\begin{equation}
V(\mathbf r)
=
V_1(\mathbf r-\mathbf R_1)+V_2(\mathbf r-\mathbf R_2).
\label{fullpotential}
\end{equation}
After scattering of DM with the full system, the exact outgoing state obeys the Lippmann--Schwinger equation~\cite{Lippmann:1950zz},
\begin{equation}
|\psi^{(+)}\rangle
=
|\mathbf k\rangle + G_0^{(+)}V|\psi^{(+)}\rangle
\end{equation}
with free Green operator defined through the free Hamiltonian without interactions $H_0$,
\begin{equation}
    G_0^{(+)}\equiv\frac{1}{E_k-H_0+i0}.
\end{equation}

According to the Ehrenfest theorem~\cite{Ehrenfest:1927swx}, the DM-induced force on target $1$ is,
${\bf F}_1 = -\braket{\nabla_{\mathbf{R}_1} V_1} = \braket{\nabla_{\mathbf{r}} V_1}$.
The expectation value of a localized gradient vanishes for plane waves. Therefore, the leading nonzero term of force is
\begin{equation}
\mathbf F_1
=
2\,\mathrm{Re}\,
\langle \mathbf k|
\nabla_\mathbf{r} V_1(\mathbf r-\mathbf R_1)\,
G_0^{(+)}V
|\mathbf k\rangle,
\label{force1}
\end{equation}
To compute the Green's operator explicitly, we insert the complete set of continuum momentum eigenstates into the above \geqn{force1} as
\begin{equation}
    \mathbf F_1 = 2\,\mathrm{Re}\int \frac{d^3{\bf p}}{(2\pi)^3}
\frac{i(\mathbf k-\mathbf p)\,A_1(\mathbf k, \mathbf p)}{E_k-E_p+i0}
\label{force2}
\end{equation}
with a definition
\begin{equation}
A_1(\mathbf k, \mathbf p)
\equiv
\langle \mathbf k|V_1(\mathbf r-\mathbf R_1)|\mathbf p\rangle
\langle \mathbf p|V|\mathbf k\rangle
\end{equation}
for simplicity and $E_p$ is the free energy for eigenstate $\ket{{\bf p}}$.
Here, we apply the integration by parts for
$\braket{{\bf k} | \nabla V_1|{\bf p}} = i ({\bf k} - {\bf p}) \langle \mathbf k|V_1|\mathbf p\rangle$.
With the translation invariance, $\langle \mathbf k|V_1 ({\bf r} - {\bf R_1})|\mathbf p\rangle = e^{i({\bf k} - {\bf p})\cdot {\bf R}_1} \langle \mathbf k|V_1 ({\bf r})|\mathbf p\rangle$, one can build the relationship with $A_1$ and the scattering amplitude under the first order Born approximation, $f_i
=
-m_\chi
\langle \mathbf{p}|V_i|\mathbf{k}\rangle /2\pi$, as
\begin{equation}
A_1(\mathbf{k}, \mathbf p)
=
\frac{4 \pi^2}{m_\chi^2} \left( |f_1|^2
+
e^{i(\mathbf k-\mathbf p)\cdot \Delta \mathbf R}\,
f_1^* f_2 \right).
\label{A_1}
\end{equation}
The separation between the two scattering centers is defined as $\Delta \mathbf R \equiv \mathbf R_2-\mathbf R_1$.

To further evaluate \geqn{force2}, we apply the Sokhotski–Plemelj formula,
\begin{equation}
\frac{1}{E_k-E_p+i0}
=
\mathcal P\frac{1}{E_k-E_p}
-i\pi\delta(E_k-E_p),
\end{equation}
and the force can be expressed as,
\begin{equation}
\begin{split}
    {\bf F}_1
    & = 2 \int \frac{d^3 {\bf p}}{(2\pi)^3} ({\bf k} - {\bf p}) \\
  & \quad \times
\left[
\pi \delta(E_k - E_p)\,\mathrm{Re}\,A_1
- \mathcal{P}\frac{\mathrm{Im}\,A_1}{E_k - E_p}
\right].
\label{force3}
\end{split}
\end{equation}
The Cauchy principal value $\mathcal{P}$ is evaluated only at the imaginary part of $A_1$. For a general scatterer with a real potential symmetric about the center-of-mass, the Born amplitudes $f_{1(2)}$ are also real. The imaginary part arises entirely from the phase factor.

While the formalism above is general,
the physical system of interest in this work is the pair of concentric cylindrical targets in MICROSCOPE, as shown in \gfig{fig:cartoon}, so that $\Delta \mathbf R = 0$ and the phase factor also drops out. Accordingly, we keep only the real part of $A_1$ which imposes the on-shell condition, $\delta (E_p - E_k) = \delta (p-k) m_\chi/k$, in \geqn{force3}.
Therefore, the momentum ${\bf p}$ has the same magnitude as the initial DM momentum ${\bf k}$ but arbitrary direction. It thus can be physically identified as the final DM momentum ${\bf k'}$.
If we focus only on the force along the incident $z$-axis, for which $(\mathbf k-\mathbf k')_z = k(1-\cos\theta)$, one obtains
\begin{align}
F_{1(2)}^z
=
j_{\rm in} k
\int d\Omega\,(1-\cos\theta)\,
f_{\rm tot} f_{1(2)},
\label{force_int}
\end{align}
Here, $f_{\rm tot} \equiv f_1 + f_2$ is the Born amplitude of DM scattering with the full potential $V$ in \geqn{fullpotential}.
The force on target $2$ can be obtained in an analogous manner and takes a highly symmetric form.
It can even be generalized to the multiple-scatterer case.

In this way, we obtain a compact formula for the force distribution among multiple targets in their scattering with the DM flux.
It shows that, in the presence of interference, the scattering cross section is no longer determined solely by $|f_1|^2$ or $|f_2|^2$ as in \geqn{force_single}. Instead, the force experienced by each target contains an interference term originating from the scattering amplitudes of other targets, leading to a redistribution of the individual forces. This distinction will be made explicit in the next section, when we calculate the signal arising from DM wave scattering off the targets in MICROSCOPE.

\prlsection{Signatures in MICROSCOPE}{.}
MICROSCOPE is designed to test the WEP by comparing the differential acceleration of two coaxial cylindrical test masses composed of different materials in the Earth's gravitational field~\cite{MICROSCOPE:2019jix}.
The experimental sensitivity to the Eötvös parameter $\eta(2,1)= 2(a_{2} - a_{1})/(a_{2} + a_{1})$ reaches $10^{-15}$~\cite{MICROSCOPE:2022doy}.

The space mission comprises two similar differential accelerometers, each containing two concentric hollow cylindrical test masses as shown in \gfig{fig:cartoon}. The sensing axis of the accelerometer aligns with the cylinder symmetry axis ($z$-axis), and the two cylinders are nested coaxially with a radial gap of $\approx 1\;\mathrm{cm}$. The two differential accelerometers differ mainly in their compositions: in SUREF, both cylinders are made of a Pt--Rh (90:10) alloy, while in SUEP the inner cylinder is made of Pt--Rh and the outer of TA6V (Ti--Al--V, 90:6:4). The detailed dimensions and density parameters of these cylinders are summarized in Table~\ref{tab:test_mass_dimensions}~\cite{MICROSCOPE:2019jix,MICROSCOPE:2022doy}.

The spacecraft operates in a Sun-synchronous orbit at an altitude of $710~\mathrm{km}$, with an orbital inclination of $98^\circ$ relative to the Earth's equatorial plane~\cite{MICROSCOPE:2019jix,MICROSCOPE:2022doy}. During scientific session, it spins about an axis normal to the orbital plane with a frequency of $0.757 \times 10^{-3}~\mathrm{Hz}$ in V2 mode and $2.943 \times 10^{-3}~\mathrm{Hz}$ in V3 mode~\cite{MICROSCOPE:2022doy}. As a result, the DM incidence relative to the sensing axis is modulated over time, which facilitates high-precision measurements by enabling clear discrimination between the signal and background disturbances. The DM incident angle $\theta_{\rm in}$ relative to the sensing axis can be determined from the spacecraft’s attitude data referenced to the J2000 frame (for more details, see Supplemental Material).

\begin{table*}[t]
\caption{Parameters of the cylindrical test masses in both the SUEP and SUREF sensor units.~\cite{MICROSCOPE:2019jix,MICROSCOPE:2022doy}}
\label{tab:test_mass_dimensions}
\begin{ruledtabular}
\begin{tabular}{lccccc}
 & Inner Radius (mm) & Outer Radius (mm) & Length (mm) & Density ($\text{g/cm}^3$) & Mass (kg) \\
\hline
$\text{Inner Cylinder(SUEP)}$ & 15.4 & 19.7 & 43.3 & 20.0 & 0.40 \\
$\text{Outer Cylinder(SUEP)}$ & 30.4 & 34.7 & 79.8 & 4.4  & 0.30 \\
$\text{Inner Cylinder(SUREF)}$ & 15.4 & 19.7 & 43.3 & 20.0 & 0.40 \\
$\text{Outer Cylinder(SUREF)}$ & 30.4 & 34.7 & 79.8 & 20.0  & 1.36 \\
\end{tabular}
\end{ruledtabular}
\end{table*}

When the DM wave wind is incident on the two cylinders from a given direction ($\theta_{\rm in}$), the cylinders experience different forces due to the differing scattering amplitudes, resulting in an observable differential acceleration along the sensing axis:
\begin{equation}
    \Delta a_z \equiv a_z^1 - a_z^2 \equiv \frac{F_z^1}{m_1} - \frac{F_z^2}{m_2}.
    \label{da}
\end{equation}

To calculate the force $F_z$ using \geqn{force_int}, we first need to determine the DM wind flux incident on the test masses. The microscopic velocity of the DM particles relative to the spacecraft (i.e. the cylinder test masses) is given by ${\bf v}_\chi = {\bf v} -  {\bf v}_\mathrm{S}$, where ${\bf v}_\mathrm{S}$ is the spacecraft's velocity relative to the Galactic halo, which is dominated by the velocity of the Solar System in the Galaxy ($v_0 \sim 220\,\mathrm{km/s}$). The DM velocity $\mathbf v$ in the Galactic frame follows an isotropic Maxwellian distribution~\cite{Evans:2018bqy,Baxter:2021pqo}:
\begin{equation}
f_\chi (\mathbf{v}) \propto
\exp\!\left( - \frac{|\mathbf{v}|^2}{v_0^2} \right)
\Theta\bigl(v_{\mathrm{esc}} - |\mathbf{v}|\bigr),
\label{eq:dist}
\end{equation}
where $v_{\mathrm{esc}}\sim 544\,\mathrm{km/s}$ is the Galactic escape speed, and \(\Theta\) is the Heaviside step function.

For the scattering of an incident DM particle with velocity $\mathbf{v}_\chi = |\mathbf{v}_\chi|(\sin\theta_1\cos\phi_1,\; \sin\theta_1\sin\phi_1,\; \cos\theta_1)$ into an outgoing solid angle $\Omega_2$, the momentum transfer projected along the sensing axis is $q_z = m_\chi v_\chi \,(\cos\theta_1 - \cos\theta_2)$. Consequently, the net force along the $z$-axis exerted by the DM wind on a macroscopic target is given by
\begin{equation}
F_z^i(\theta_{\rm in}) = \frac{\rho_\chi}{m_\chi} \int d^3\mathbf{v}_\chi \, f_\chi (\mathbf{v}_\chi + \mathbf{v}_\mathrm{S})\, v_\chi
\int d\Omega_2 \, f_{\rm tot} f_i \,
q_z,
\label{eq:Fz_full}
\end{equation}
where $\theta_{\rm in}$ is the polar angle of $\mathbf{v}_\mathrm{S}$ relative to the sensing axis.

\begin{figure}[t]
    \centering
    \includegraphics[width=0.9\linewidth]{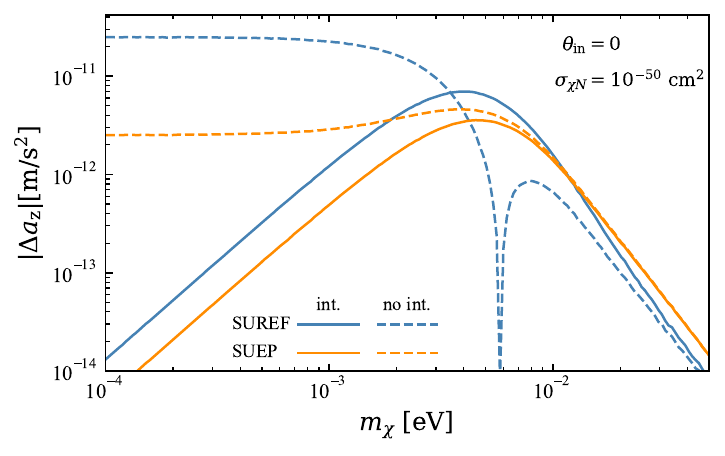}
    \caption{DM-scattering-induced accelerations on SUEP (yellow) and SUREF (blue) with (solid, int) and without (dashed, no int) quantum interference effects included. $\theta_\mathrm{in} = 0$ and $\sigma_{\chi N} = 10^{-50}\,$cm$^2$ are used for calculations.}
    \label{fig:adiff}
\end{figure}

To illustrate the significance of quantum interference, Fig.~\ref{fig:adiff} compares the differential accelerations for SUREF and SUEP calculated with interference using \geqn{eq:Fz_full} and without interference by replacing $f_{\rm tot}$ with $f_i$ in \geqn{eq:Fz_full}.
For large DM masses with short wavelengths, the two calculations nearly coincide, as the DM wave predominantly probes local nucleons and remains insensitive to macroscopic interference.
As the DM wavelength increases, particularly when it far exceeds the radial gap between the cylinders, the discrepancy becomes increasingly pronounced, underscoring the necessity of accounting for quantum interference to ensure accurate results. 
Specifically, in the limit where the wavelength is much larger than the cylinder dimensions, a calculation neglecting interference would imply that DM scatters coherently off all nucleons within each individual target. Consequently, the more massive target would experience a larger force, with the acceleration difference asymptotically approaching a constant. In contrast, when interference is properly included, the acceleration difference vanishes at long wavelengths because the two targets become indistinguishable as a whole from the perspective of the DM wave. The dip in the dashed curves originates from a coincidental cancellation in the non-interfering case, where the two cylinders happen to acquire identical accelerations. 

\begin{figure}[t]
    \centering
    \includegraphics[width=0.9\linewidth]{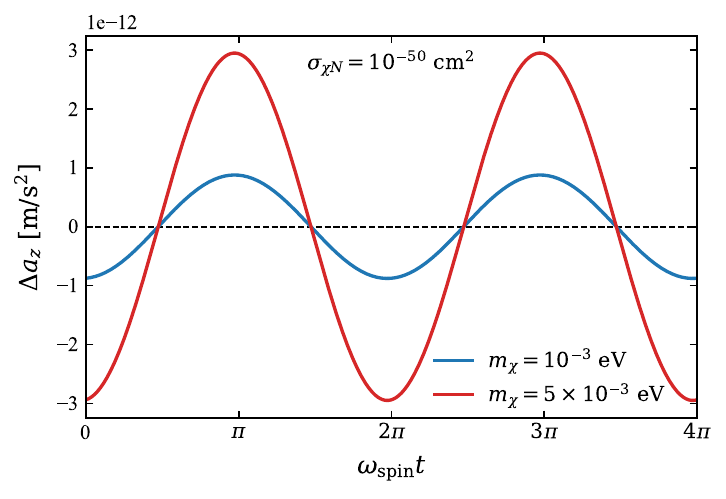}
    \caption{DM-induced differential acceleration $\Delta a_z$ for SUEP versus spin phase. Calculations are based on MICROSCOPE session 218 attitude data. The curves correspond to $m_\chi=10^{-3}\,\mathrm{eV}$ (blue) and $5\times10^{-3}\,\mathrm{eV}$ (red), respectively; the cross section is $\sigma_{\chi N}=10^{-50} \mathrm{cm}^2$.}
    \label{fig:timemod}
\end{figure}

Fig.~\ref{fig:timemod} shows a representative DM-induced differential acceleration $\Delta a_z$ as a function of the spin phase $\omega_{\rm spin}t$, computed using the MICROSCOPE session 218 attitude data and averaged over the DM velocity distribution. The signal is dominated by the spin-synchronous harmonic and exhibits an approximately sinusoidal modulation, arising from the projection of the DM induced acceleration onto the rotating sensitive axis. The modulation amplitudes for different masses are consistent with the behavior shown in  Fig.~\ref{fig:adiff}.

\prlsection{Limits on DM-nucleon Scattering}{.} As established above, the DM wave flux scatters coherently off the MICROSCOPE test masses, inducing a differential acceleration modulated by the satellite's spin. Incorporating quantum interference effects is essential for obtaining correct results, as they redistribute forces between the two test masses. With this formalism, we can then extract the DM–nucleon scattering cross section by analyzing the differential acceleration at the spin frequency within the MICROSCOPE experimental data.

We adopt an analysis procedure analogous to that used for determining the Eötvös parameter in the MICROSCOPE mission~\cite{MICROSCOPE:2022doy}. The N2B-level differential acceleration ($\Gamma_{z,\mathrm{corr}}^{(d)}$), corrected for calibrated parameters and gravity gradient effect, is fitted to the following signal model:
\begin{equation}
\Gamma_{z,\mathrm{corr}}^{(d)}(t)
=
\sum_{j=0}^{3}\alpha_j (t-t_0)^j
+
\Delta a_{z}(m_\chi,t;\sigma_{\chi N})
+
n_z^{(d)}(t).
\label{eq:ac_fit}
\end{equation}
Here, the summation term is a cubic polynomial accounting for low-frequency drifts, $n_z^{(d)}$ denotes the random noise, and the fit is used to extract $\sigma_{\chi N}$ for a given DM mass $m_\chi$ (see Supplemental Material for more details). Taking $m_\chi = 5\times10^{-3}\ \mathrm{eV}$ for example, the combined results across all sessions yield $\sigma_{\chi N} = (-5.5\pm1.6_{\rm stat}\pm4.2_{\rm sys})\times10^{-53}\ \mathrm{cm^2}$ for SUREF and $\sigma_{\chi N} =(14.9\pm5.3_{\rm stat}\pm4.5_{\rm sys})\times10^{-53}\ \mathrm{cm^2}$ for SUEP, with uncertainties quoted at the 1$\,\sigma$ confidence level. Given the proximity of the DM modulation frequency to the EP violation frequency, we approximate the systematic uncertainty using the value determined at the EP violation frequency~\cite{rodrigues2022microscope}. These results show no significant evidence of a DM signal; consequently, constraints on $\sigma_{\chi N}$ are established at the 2$\,\sigma$ confidence level.

\begin{figure}[t]
    \centering
    \includegraphics[width=0.9\linewidth]{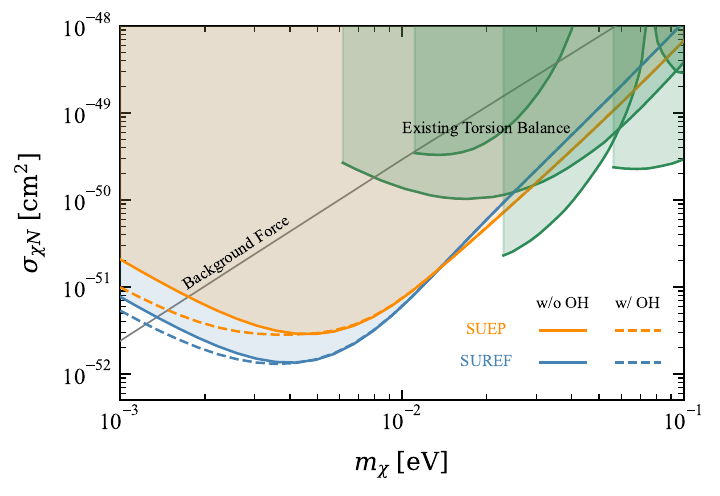}
    \caption{Constraints on the DM-nucleon scattering cross section as a function of DM mass. The yellow and blue lines represent limits derived from the SUEP and SUREF sensor units, respectively. Dashed and solid lines indicate the results with and without including the quantum interference effects from the sensor units' outer housing (OH). For comparison, existing constraints from torsion balance experiments (green region) and background forces (gray solid line) are also shown.}
    \label{fig:limits}
\end{figure}

The resulting constraints for the SUEP (yellow) and SUREF (blue) sensor units are shown as solid lines in Fig.~\ref{fig:limits}.
Consistent with the behavior observed in  Fig.~\ref{fig:adiff}, the constraints weaken for heavier DM with shorter wavelengths because coherent scattering is suppressed. In the long-wavelength regime, the weakening of the constraints is a consequence of interference effects. The most stringent constraint of approximately $10^{-52}\,$cm$^2$ is obtained at $5\times10^{-3}\,$eV, where the DM coherence length, $1/(m_\chi v_\chi)$, is comparable to the sizes of the cylinders.

In comparison, ground-based torsion-balance experiments provide stronger constraints in the higher-mass region, owing to the smaller physical dimensions of their test masses ~\cite{Matsumoto:2025rcz}.
Additionally, quadratically coupled DM can induce mediated forces between nucleons~\cite{VanTilburg:2024xib, Barbosa:2024pkl, Grossman:2025cov, Cheng:2025fak, Gan:2025nlu}. This background-force constraint ~\cite{VanTilburg:2024xib} from short-range EP tests~\cite{Tan:2020vpf} is shown as the gray curve.
Although the DM coupling in the parameter space considered here is too small for DM to thermalize during BBN and thereby affect the relativistic degrees of freedom, it can still modify the effective nucleon masses, indirectly affecting BBN observables. However, the resulting constraint is weaker than the background-force constraint~\cite{Bouley:2022eer}.
Notably, in the mass range $10^{-3}$--$10^{-2}$\,eV, MICROSCOPE provides the most stringent constraints on the DM–nucleon scattering cross section, improving upon existing bounds by two orders of magnitude.

We also account for the contribution of the outer housing (OH) surrounding the two test masses to the DM-induced signal. The OH is modeled as a concentric cylindrical shell with a mass of $0.935$\,kg. The dashed curves represent the resulting constraints when considering coherent DM scattering off all three bodies. Since the force signal in \geqn{eq:Fz_full} is proportional to the total scattering amplitude $f_{\rm tot}$, it is worth noting that the inclusion of the OH as an additional scatterer enhances the overall signal, thereby yielding more stringent constraints. This effect provides guidance for designing interference-enhanced detection of DM in the future.

\prlsection{Conclusion}{.} We have established the framework that coherent scattering of ultralight DM from multiple macroscopic targets leads to a force-distribution effect absent for isolated scatterers. When the target sizes and separations are comparable to the DM wavelength, the scattering amplitudes from different bodies interfere, producing distinctive differential-force signals.

We applied this general framework to the nested cylindrical test masses of MICROSCOPE. The satellite spin converts the direction-dependent response into a time-modulated acceleration signal.
By analyzing this modulation and the MICROSCOPE sensitivity at the corresponding frequency, we obtain the strongest constraints on DM--nucleon elastic scattering in the mass range $10^{-3}$--$10^{-2}\,$eV, reaching cross sections of order $10^{-52}\,$cm$^2$. These results establish multi-body precision accelerometers as probes of wave-interference signatures in DM scattering.

\section*{Acknowledgements}

C.-T. F. and R. L. are supported by the National Key R\&D Program of China (Grant No.~2024YFC2207500).
P.-S. L. is supported by the National Key R\&D Program of China (Grant No.~2022YFC2204100) and the National Natural Science Foundation of China (NSFC)(Grant No.~12475052).
J. S. is supported by the Japan Society for the Promotion of Science (JSPS) as a part of the Postdoctoral Program (Standard) with grant number P25018 and
by the World Premier International Research Center Initiative (WPI), MEXT, Japan (Kavli IPMU).
C.-Y. X. is supported by the Fundamental Research Funds for the Central Universities (No.~24CX06048A).

\section*{Data availability}

The MICROSCOPE data products used in this work are publicly available from the MICROSCOPE data center. These data are ``Products acquired by CNES-ONERA-CNRS-OCA-ESA-DLR-ZARM MICROSCOPE Mission''.

\clearpage

\begin{center}
{\large \bf Supplemental Material for\\
``Macroscopic Quantum Interference in Dark Matter Wave Scattering \\ with MICROSCOPE''}
\end{center}

This Supplemental Material describes how the dark-matter (DM) and nucleon
scattering cross section $\sigma_{\chi N}$ is extracted from the MICROSCOPE differential
acceleration data.  For each MICROSCOPE scientific session, we fit the corrected differential
acceleration with a DM signal template, estimate the statistical and
systematic uncertainties, and then combine the results from different
sessions.

\noindent\textbf{Fitting model.} For the $l$-th MICROSCOPE session and a DM mass $m_\chi$, the
corrected differential acceleration provided by the N2B level data is modeled as
\begin{equation}
\begin{aligned}
\Gamma_{z,{\rm corr}}^{(d)}(t)
&=
p_l(t)
+
\sigma_{\chi N}(m_\chi)
\Delta \tilde a_{z,l}(m_\chi,t)
+
n_{z,l}^{(d)}(t),
\\
p_l(t)
&=
\sum_{j=0}^{3}
\alpha_{j,l}(t-t_{0,l})^j .
\end{aligned}
\label{eq:time_model_appendix}
\end{equation}
Here $\Gamma_{z,{\rm corr}}^{(d)}$ is the corrected differential
acceleration along the sensitive axis, $p_l$ is a cubic polynomial used
to account for residual low-frequency drifts, and
$n_{z,l}^{(d)}$ denotes the residual noise.  The quantity
$\Delta \tilde a_{z,l}(m_\chi,t_n)$ is the session-dependent DM-induced
differential acceleration template per unit DM--nucleon scattering cross
section.  The physical DM-induced differential
acceleration is
\begin{equation}
\Delta a_{z,l}(m_\chi,t;\sigma_{\chi N})
=
\sigma_{\chi N}(m_\chi)
\Delta \tilde a_{z,l}(m_\chi,t),
\label{eq:unit_cross_section_template}
\end{equation}
where the scattering cross section $\sigma_{\chi N}$ is the fitting parameter.

\noindent\textbf{Frequency-domain Least Squares Fitting.} Following the MICROSCOPE data analysis strategy\cite{touboul2022result}, the fit is performed in the
frequency domain within a narrow band around the expected modulation
frequency.  For the DM wind signal considered here, the dominant modulation occurs
at the satellite spin frequency.  We therefore fit the data in a narrow
band centered at \(f_{{\rm spin},l}\),
\begin{equation}
\mathcal B_l
=
\left\{
f_i:
|f_i-f_{{\rm spin},l}|
\leq
{\Delta f_l\over 2}
\right\}.
\end{equation}

In the frequency domain, Eq.~(\ref{eq:time_model_appendix}) can be
written as
\begin{equation}
Y_l(f_i)
=
P_l(f_i)
+
\sigma_{\chi N}(m_\chi)
S_l(f_i;m_\chi)
+
N_l(f_i),
\label{eq:frequency_model_appendix}
\end{equation}
where $Y_l(f_i)$, $P_l(f_i)$, $S_l(f_i;m_\chi)$, and $N_l(f_i)$ are the
Fourier components of the corrected acceleration, the polynomial drift,
the unit-cross-section DM signal, and the residual noise, respectively.
For each $m_\chi$, the frequency-domain weighted least-squares fit
gives the session-level estimator $\hat\sigma_{\chi N,l}(m_\chi)$ and
its statistical uncertainty
$\Delta\sigma_{\chi N,l}^{\rm stat}(m_\chi)$.
The frequency weights are set by the noise power spectral density
estimated in the same fitting band,
as in the MICROSCOPE frequency-domain analysis.

\noindent\textbf{Systematic uncertainty.} Since the dominant DM modulation frequency is close to that of the Equivalence Principle (EP) violation signal, we use the systematic uncertainty evaluated at the EP violation frequency as an approximation for the DM case. Consequently, the systematic uncertainty of the differential acceleration is given by
\begin{equation}
\sigma_{A,l}^{\rm sys}({\rm DM})
=
\sigma_{A,l}^{\rm sys}({\rm EP}) .
\label{eq:sys_dm_ep_appendix}
\end{equation}
To convert this acceleration uncertainty into an uncertainty on
$\sigma_{\chi N}$, we use the Fourier amplitude of the unit-cross-section
DM signal at the spin frequency. Therefore
the systematic uncertainty on the
fitted cross section is therefore given by
\begin{equation}
\sigma_{\chi N,l}^{\rm sys}(m_\chi)
=
{
\sigma_{A,l}^{\rm sys}({\rm DM})
\over
\left|
S_{l}(f_{{\rm spin},l},m_\chi)
\right|
}.
\label{eq:sigma_sys_appendix}
\end{equation}

The session-level fitted cross sections for $m_\chi=5\times10^{-3}\,\mathrm{eV}$
without the outer housing are summarized in
Tables~\ref{tabsuref_wooh_m5e3} and \ref{tabsuep_wooh_m5e3}.

\begin{table*}[htbp]
\centering
\small
\caption{Session-level fitted cross section for SUREF w/o OH at
$m_\chi=5\times10^{-3}\,\mathrm{eV}$.}
\label{tabsuref_wooh_m5e3}
\begin{tabular*}{0.78\textwidth}{@{\extracolsep{\fill}}lccc}
\hline
Session &
$\hat{\sigma}_{\chi N}\,[10^{-52}\,\mathrm{cm^2}]$ &
$\Delta\sigma_{\chi N}^{\rm stat}\,[10^{-52}\,\mathrm{cm^2}]$ &
$\Delta\sigma_{\chi N}^{\rm sys}\,[10^{-52}\,\mathrm{cm^2}]$ \\
\hline
120-1 & $-1.68$ & $2.24$ & $1.05$ \\
120-2 & $0.83$ & $1.13$ & $1.05$ \\
174 & $-0.07$ & $0.69$ & $0.91$ \\
176 & $0.23$ & $0.67$ & $1.09$ \\
294 & $0.87$ & $0.40$ & $0.49$ \\
376-1 & $1.15$ & $1.02$ & $0.71$ \\
376-2 & $0.52$ & $0.72$ & $0.71$ \\
380-1 & $-1.88$ & $0.33$ & $0.16$ \\
380-2 & $-2.37$ & $0.30$ & $0.16$ \\
452 & $1.57$ & $0.97$ & $0.49$ \\
454 & $0.35$ & $0.53$ & $0.63$ \\
778-1 & $-0.82$ & $0.73$ & $0.64$ \\
778-2 & $0.40$ & $0.83$ & $0.64$ \\
\hline
\end{tabular*}
\end{table*}

\begin{table*}[htbp]
\centering
\small
\caption{Session-level fitted cross section for SUEP w/o OH at
$m_\chi=5\times10^{-3}\,\mathrm{eV}$.}
\label{tabsuep_wooh_m5e3}
\begin{tabular*}{0.78\textwidth}{@{\extracolsep{\fill}}lccc}
\hline
Session &
$\hat{\sigma}_{\chi N}\,[10^{-52}\,\mathrm{cm^2}]$ &
$\Delta\sigma_{\chi N}^{\rm stat}\,[10^{-52}\,\mathrm{cm^2}]$ &
$\Delta\sigma_{\chi N}^{\rm sys}\,[10^{-52}\,\mathrm{cm^2}]$ \\
\hline
210 & $11.78$ & $3.80$ & $0.77$ \\
212 & $1.71$ & $3.44$ & $0.43$ \\
218 & $-1.90$ & $2.27$ & $0.46$ \\
234 & $1.89$ & $2.13$ & $0.39$ \\
236 & $2.69$ & $1.53$ & $0.46$ \\
238 & $0.22$ & $1.42$ & $0.45$ \\
252 & $1.70$ & $1.76$ & $0.39$ \\
254 & $1.66$ & $1.46$ & $0.53$ \\
256 & $-0.52$ & $1.86$ & $0.39$ \\
326-1 & $3.63$ & $2.23$ & $0.65$ \\
326-2 & $-1.79$ & $4.02$ & $0.59$ \\
358 & $0.85$ & $2.36$ & $0.40$ \\
402 & $12.66$ & $10.83$ & $2.65$ \\
404 & $2.63$ & $1.54$ & $0.37$ \\
406 & $-0.35$ & $4.48$ & $1.20$ \\
438 & $-4.20$ & $7.06$ & $2.25$ \\
442 & $2.65$ & $5.95$ & $3.05$ \\
748 & $26.93$ & $17.99$ & $3.43$ \\
750 & $5.46$ & $11.16$ & $3.36$ \\
\hline
\end{tabular*}
\end{table*}

\noindent\textbf{Combination of sessions.} The results from different sessions are then combined to give a final measurement on $\sigma_{\chi N}$. The combined mean value of $\sigma_{\chi N}$ is given by
\begin{equation}
\bar\sigma_{\chi N}(m_\chi)
=
{
\sum_l
w_l(m_\chi)
\hat\sigma_{\chi N,l}(m_\chi)
\over
\sum_l
w_l(m_\chi)
},
\label{eq:sigma_comb_mean_appendix}
\end{equation}
and the combined statistical uncertainty is
\begin{equation}
\sigma_{\chi N}^{\rm stat}(m_\chi)
=
\left[
\sum_l w_l(m_\chi)
\right]^{-1/2},
\label{eq:sigma_comb_stat_appendix}
\end{equation}
where the statistical weights is give by
\begin{equation}
w_l(m_\chi)
=
\left[
\sigma_{\chi N,l}^{\rm stat}(m_\chi)
\right]^{-2}.
\label{eq:stat_weights_appendix}
\end{equation}
The systematic uncertainty is combined using the same statistical
weights as
\begin{equation}
\sigma_{\chi N}^{\rm sys}(m_\chi)
=
{
\sum_l
w_l(m_\chi)
\sigma_{\chi N,l}^{\rm sys}(m_\chi)
\over
\sum_l
w_l(m_\chi)
}.
\label{eq:sigma_comb_sys_appendix}
\end{equation}
Finally, the conservative two-standard-deviation upper limit is defined as
\begin{equation}
\begin{aligned}
\sigma_{\chi N}^{\rm lim}(m_\chi)
&=
\left|
\bar\sigma_{\chi N}(m_\chi)
\right|
\\
&\quad
+
2
\left\{
\left[
\sigma_{\chi N}^{\rm stat}(m_\chi)
\right]^2
+
\left[
\sigma_{\chi N}^{\rm sys}(m_\chi)
\right]^2
\right\}^{1/2}.
\end{aligned}
\label{eq:sigma_limit_appendix}
\end{equation}

\noindent\textbf{Dark Matter Signal Template construction.} We now describe the construction of the session-dependent template $\Delta \tilde a_{z,l}(m_\chi,t_n)$.  As established in the main text, the DM-induced
force—and consequently $\Delta \tilde a_{z,l}(m_\chi,t_n)$—on a macroscopic target is written as a function of the incidence angle $\theta_{\rm in}$. This angle is defined as the polar angle of the velocity
$\mathbf v_\mathrm{S}$ relative to the sensing axis.  For each MICROSCOPE session, we determine $\theta_{\rm in}(t_n)$ using the satellite attitude data.

In the J2000 frame, the apparent DM-wind velocity is \cite{mccabe2014earth}
\begin{equation}
\mathbf v_\mathrm{S}^{\rm J2000}(t)
=
\mathbf V_{\odot}^{\rm J2000}
+
\mathbf V_{\oplus}^{\rm J2000}(t),
\label{eq:vS_j2000_appendix}
\end{equation}
where $\mathbf V_{\odot}^{\rm J2000}$ is the velocity of the Sun with
respect to the Galactic halo, and
$\mathbf V_{\oplus}^{\rm J2000}(t)$ is the Earth's orbital velocity around the Sun.  The
satellite orbital velocity is neglected in this step because it is much
smaller than the characteristic halo speed.

The attitude quaternions provided in the N0 level data define the rotation matrix $\mathbf R_{{\rm sat}\leftarrow{\rm J2000}}$, which transforms the velocity from the J2000 frame to the satellite body frame as follows:
\begin{equation}
\mathbf v_\mathrm{S}^{\rm sat}(t)
=
\mathbf R_{{\rm sat}\leftarrow{\rm J2000}}(t)
\mathbf v_\mathrm{S}^{\rm J2000}(t).
\label{eq:vS_sat_appendix}
\end{equation}
 The sensor frame is related to the satellite body frame via the following transformation
\begin{equation}
\begin{aligned}
\mathbf v_\mathrm{S}^{\rm sens}(t)
&=
\mathbf C_{{\rm sens}\leftarrow{\rm sat}}
\mathbf v_\mathrm{S}^{\rm sat}(t),
\\
\mathbf C_{{\rm sens}\leftarrow{\rm sat}}
&=
{\rm diag}(1,-1,-1).
\end{aligned}
\label{eq:sensor_rotation_appendix}
\end{equation}
The incidence angle of the dark matter wind relative to the sensing axis is then given by
\begin{equation}
\begin{aligned}
\theta_{\rm in}(t)
&=
\arccos
\left[
\hat{\mathbf v}_\mathrm{S}^{\rm sens}(t)
\cdot
\hat{\mathbf z}_{\rm sens}
\right],
\\
\hat{\mathbf v}_\mathrm{S}^{\rm sens}(t)
&\equiv
{
\mathbf v_\mathrm{S}^{\rm sens}(t)
\over
|\mathbf v_\mathrm{S}^{\rm sens}(t)|
}.
\end{aligned}
\label{eq:theta_in_appendix}
\end{equation}

For each $m_\chi$, the DM-induced differential acceleration is first computed over an array of $\theta_{\rm in}$ assuming a unit DM--nucleon scattering cross section. For each session, the actual DM-induced acceleration is then obtained by interpolating the pre-calculated data to the attitude-derived angle. By utilizing the real MICROSCOPE attitude data, this procedure preserves the true phase and modulation pattern of the DM signal in each session.

\bibliographystyle{utphys}
\bibliography{ref}

\end{document}